**Short-range Atomic Topology of *Ab initio* Generated Amorphous PdSi Alloys.**

Isaías Rodríguez[1], Renela M. Valladares[2], Alexander Valladares[2], David Hinojosa-Romero[1] and Ariel A. Valladares[1] *

[1] Instituto de Investigaciones en Materiales, Universidad Nacional Autónoma de México, Apartado Postal 70-360, Ciudad Universitaria, CDMX, 04510, México.

[2] Facultad de Ciencias, Universidad Nacional Autónoma de México, Apartado Postal 70-542, Ciudad Universitaria, CDMX, 04510, México

*Corresponding author; e-mail: valladar@unam.mx

**Abstract**

**Since the pioneering efforts of Duwez and coworkers in 1965, when a solid amorphous phase of palladium-silicon (*a*-PdSi) was obtained in the vicinity of the eutectic concentration, much work has been done. However, some points related to the atomic structures remain to be systematized. In this work, 8 amorphous $Pd_{100-c}Si_c$ alloys (c = 2.5, 5, 10, 13.34, 15, 17.5, 20, and 22 at %) were generated by molecular dynamics *ab initio* simulations; the short-range structure is analyzed using several correlation functions, like Pair Distribution Functions, reduced Pair Distribution Functions, Plane Angle Distribution Functions. Other related properties, like nearest-neighbors and some Frank-Kasper polyhedra are reported. The generated samples correctly reproduce the scarce experimental pair correlation functions that are reported in the literature. An unexpected outcome is the appearance of structural changes in the neighborhood of $Pd_{86.66}Si_{13.34}$, which may be related to the magnetic changes reported in the liquid and amorphous Pd-Si alloys previously reported. The explicit amorphous topologies are reported.**

**Keywords:** Molecular Dynamics, Magnetic materials, Amorphous materials, First Principles simulation, Radial Distribution Functions, Plane-Angle Distributions

**Introduction**

As part of a study to investigate the possible magnetic properties of the palladium-rich region of the amorphous palladium-silicon system, *a*-PdSi [1], a detailed analysis of the amorphous structures generated in the process was undertaken, structures that were systematically developed to make possible a study of the evolution of their topology using consistent simulational parameters. Evidently, it is necessary to create amorphous structures throughout the whole concentration range to produce a complete characterization of the PdSi system; however, since the predicted magnetism for the solid alloys disappeared for concentrations of the order of 20 atomic percent, at %, silicon [1], the study of higher concentrations of silicon were not carried out, but a complementary, more extensive study, is a must.

Palladium is a peculiar elemental solid since its location in the periodic table places it in a transitional state from partially filled d-band solids to completely full d-band atomic structures. A metal with paramagnetism fully corroborated in the crystalline state, experiences the promotion of electrons from the full d-band to the energetically near s-state due to the closeness of the energy levels involved. Palladium has also manifested magnetic properties of the ferro-type when expanded [2], when amorphized [3] and when contaminated with hydrogen, deuterium or tritium [4] or with Si [1].

On the other hand, silicon is a semiconductor in the condensed crystalline and amorphous states with an energy gap of the order of 1 eV. Alloying these two elements would certainly generate unexpected properties due to the dissimilar nature of the constituents. One obvious characteristic would be the transition from the metallic character of pure Pd, to the semiconductive nature of pure Si, determined by the appearance of an electronic gap at some concentration of silicon as the metal evolves into a semiconductor. The atomic structure is another characteristic that would change drastically from face-centered cubic to the diamond-like topology for the semiconductor. Evidently, these atomic arrangements would be difficult to assess for the amorphous, whereas it should be an observable behavior for the crystalline. However, the appearance of an electronic energy gap should be observable when silicon is approached, if we calculate the electronic density of sates, eDoS, and study its evolution as a function of concentration. A third characteristic would be the transition of the chemical bond from metallic to covalent as the silicon concentration increases.

Amorphous PdSi alloys became one of the first Bulk Metallic Glasses, BMG, back in 1965 and remains one of the simplest [5]. Looking at the literature [6], these alloys participate in more complicated combinations in some BMG and thereby they are an important system to study. Some recent work has been reported in an effort to produce amorphous pure Pd [7] since this phase has eluded the experimentalist, whereas an amorphous silicon sample is not as difficult to accomplish. It has been argued that by reducing the size of the system to nanoscales it is possible to generate a sample of amorphous Pd. However, when one looks at the shape of the reduced pair distribution function reported in Ref 7, one has to assume this accomplishment with care since the overall shape of this distribution function seems to resemble that of a liquid more than that of an amorphous. The synthesizing path is interesting and may eventually lead to creating a sample of *a*-Pd on a larger scale. Hydrogen absorption experiments conducted on these nanoscale specimens showed that the Pd metallic glass generated expanded little upon hydrogenation compared to the crystalline samples.

In this work, the study of 8 amorphous $Pd_{100-c}Si_c$ alloys (c = 2.5, 5, 10, 13.34, 15, 17.5, 20, and 22 at %) is reported; all specimens were generated by means of *ab initio* molecular dynamics simulations. The short-range structure is analyzed using several correlation functions: Pair Distribution Functions, PDFs, reduced Pair Distribution Functions, rPDFs, Radial Distribution Functions, RDFs, and Plane Angle Distribution Functions, PADs [Chapter 30, pp 1045-1050 of 8]. Some Coordination Numbers, CNs, and some Frank-Kasper, FK, polyhedra are also reported. The results for the pure amorphous palladium samples are taken from Ref. 3.

**Method**

The amorphous PdSi samples were obtained using the *undermelt-quench* approach that evolved in our group and that generates very good amorphous specimens [9,10] when compared to experiments. In this approach the starting point is an unstable crystalline structure which is subjected to a thermal Molecular Dynamics, MD, procedure from 300 K to a point below the liquidus (*undermelt*) to avoid melting (inhibiting structures that are reminiscent of the liquid) in 100 simulational steps. Starting from an unstable atomic arrangement fosters the disordering process and makes amorphization feasible. After reaching the highest temperature below the liquid, a *quenching* procedure is executed down to the proximity of 0 K with the same (in absolute value) thermal slope as the heating ramp.

The highest temperature depends on the phase diagram of the system and for the case of the PdSi alloys, the highest temperature was 1500 K for concentrations c < 10 at %, and 1000 K for c > 10 at % [1]. At the end of this process, the disorder has been attained and a Geometry Optimization, GO, is carried out to let the disordered structure evolve into the amorphous one at a *local* energy minimum. It is evident that the amorphous arrangement is not the *minimum-minimorum* of the energy; such minimum energy structures would be the crystalline ones.

The MD and GO processes were carried out using CASTEP [11], included in the Materials Studio suite of codes [12]. CASTEP is a Density Functional Theory, DFT, code that uses a plane wave basis set for the description of the valence electrons, with ultrasoft pseudopotential for the core treatment [13]. The exchange-correlation contribution is calculated with the Generalized Gradient Approximation, GGA, as parameterized by Perdew, Burke and Ernzerhof, PBE [14]. For energy minimization, a Pulay mixing scheme was employed, with a thermal smearing of 0.1 eV for the electronic occupation.

The MD thermal treatment was carried out with an NVT ensemble and a Nosé-Hoover thermostat. These processes consist of a disordering heating ramp of 100 steps and a quenching ramp of 125 steps to 12 K for c < 10 at %, and to 7 K for c > 10 at %; each one of these steps is of 7 fs, for a total MD time of 1.575 ps, with a thermal slope of 12 K step$^{-1}$ for c < 10 at %, and to 7 K step$^{-1}$ for c > 10 at %. A 300 eV cutoff energy was used for the plane-wave basis set, together with a Self-Consistent Field, SCF, energy threshold of $2.0 \times 10^{-6}$ eV. The total spin of the specimens was not fixed during the heating and the cooling procedure, and neither during the GO processes, so the final structures were obtained with the spin unrestricted to allow for the evolution dictated by the interactions and the procedure. For the GO processes, the SCF energy threshold was $5.0 \times 10^{-7}$ eV, and these were carried out with the BFGS minimizer with the following tolerances: energy tolerance of $5.0 \times 10^{-7}$ eV, force tolerance of $1.0 \times 10^{-2}$ eV Å$^{-1}$, and a maximum displacement of $5.0 \times 10^{-4}$ Å.

All the alloys were randomly arranged in a (non-stable) diamond-like supercell containing a total of 216 atoms of both elements. The results for the pure sample of Pd were taken from Ref 3. Care was exercised to construct these supercells using the experimental densities reported in the literature [15–17]. The resulting structures were then analyzed using **Correlation,** a code developed by Rodríguez *et al.* [18], with the purpose of calculating several correlation functions and properties derived from these functions.

**Results and Discussion**

The comparison between our PDF results and the existing experimental ones can be seen in Figures 1. The agreement with Fukunaga and Suzuki is remarkable for $Pd_{78}Si_{22}$, Figure 1 a); for $Pd_{80}Si_{20}$, Figure 1 b), and for $Pd_{85}Si_{15}$, Figure 1 d) [16]. There is a very good agreement for the first peak of the Pd-Pd partial PDF, pPDF, for the Pd85Si15 with Andonov [19] but there seems to be some discrepancy in the first peak of the Pd-Si pPDF, since Andonov report a peak bellow 2Å as can be seen in Figure 1c). Meanwhile, the Pd-Si bond length reported in the literature varies from 2.1 Å to 2.5 Å [REFs 1], in agreement with this work's results of 2.409 Å on average for the Pd-Si bond length, as seen in Table 1. A comparison of the reduced Pair Distribution Function, rPDF, of Mrafko and Duhaj*.* with our results is shown in Figure 1 e) [20]. Finally, the comparison of the RDF of Crewdson with our results can be seen in Figure 1 f) [21]. The second coordination shell is indicated in detail in the insets of Figures 1.

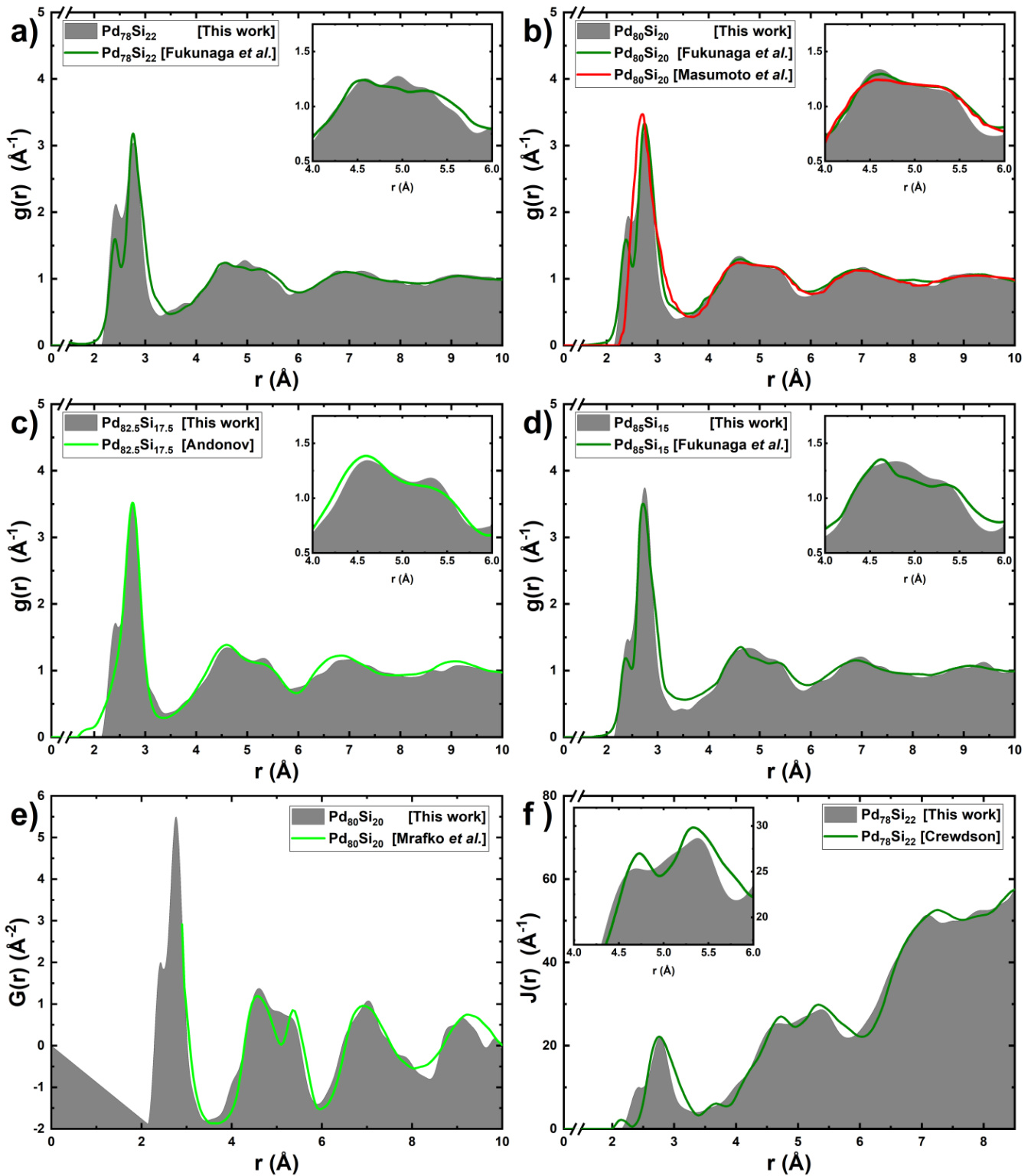

**Figure 1.** Comparison between existing experimental distribution functions and our simulations. a) $Pd_{78}Si_{22}$ PDF *vs* Fukunaga and Suzuki [13], b) $Pd_{80}Si_{20}$ PDF *vs* Fukunaga and Suzuki [16] and Masumoto as reported by Waseda [16,22], c) $Pd_{82.5}Si_{17.5}$ PDF *vs* Andonov [19], d) $Pd_{85}Si_{15}$ PDF *vs* Fukunaga and Suzuki [16], e) $Pd_{80}Si_{20}$ rPDF *vs* Mrafko and Duhaj [20], f) $Pd_{78}Si_{22}$ RDF *vs* Crewdson [21]. The insets show the second coordination shell.

In Figure 2 a) the fact that the first coordination shell of the PdSi alloys is composed of two peaks can be observed. The first peak is due to the Pd-Si partial and is located at 2.43 Å for c = 22 % and decreases down to 2.39 Å for c = 2.5 % as the Si concentration decreases, as can be seen in Figure 2 c). Fukunaga and Suzuki reported 2.42 Å for c = 22 % [16], very close to our results. The second peak of the first coordination shell is due to the partial Pd-Pd, which can be seen in Figure 2 b), this peak is located at 2.77 Å for c = 22 % and decreases as the Si concentration decreases, to 2.69 Å for pure palladium; Fukunaga and Suzuki reported 2.78 Å for c = 22 % [16]. There seems to be an absence of first neighbors in the Si-Si partial as can be seen in Figure 2 d), the first visible peak in the Si-Si partial is located around 3.9 Å, a second neighbor interaction because the Si-Si bond length is 2.36 Å for the crystalline solid.

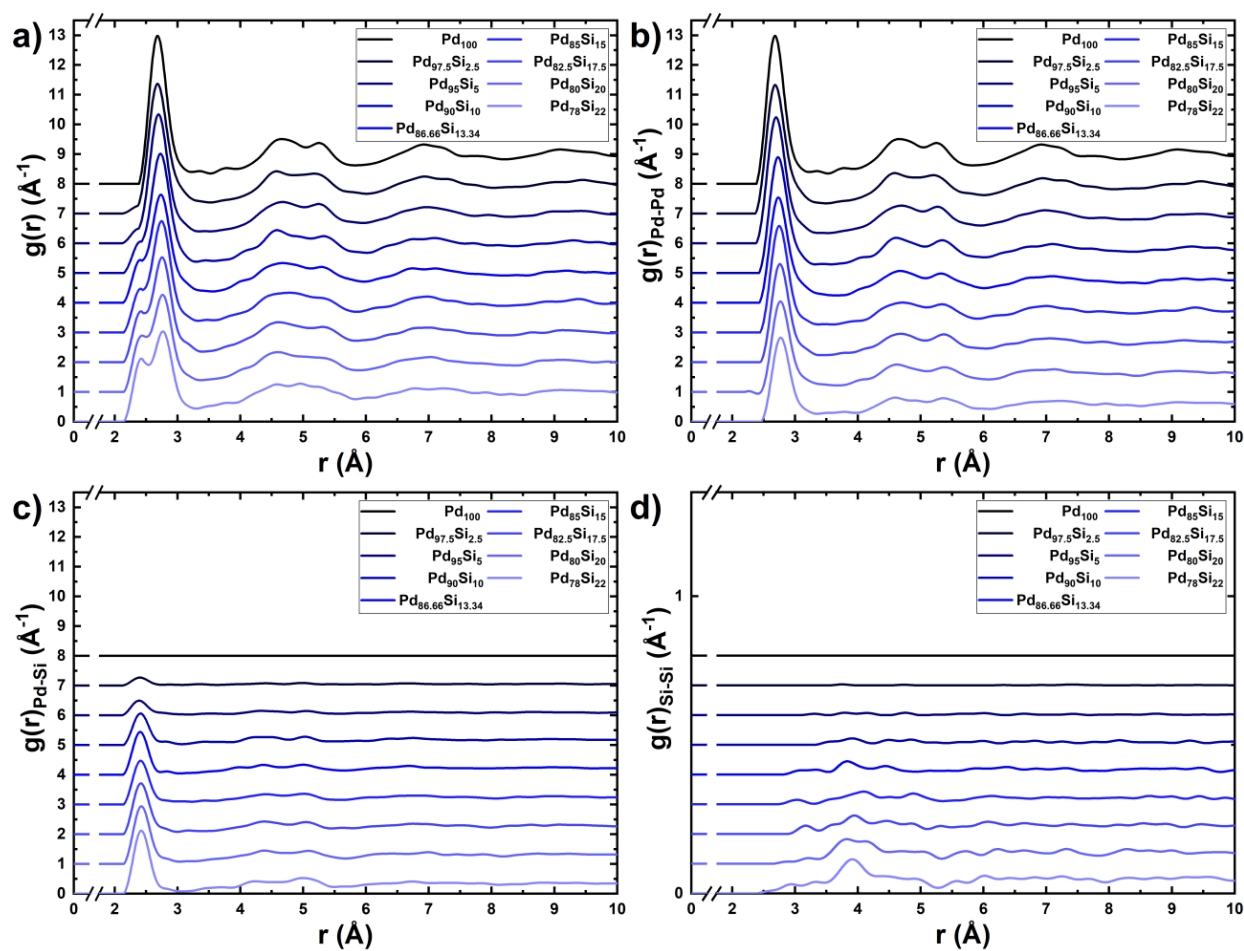

**Figure 2.** Pair Distribution Functions (PDFs) for the eight PdSi alloys studied and for the pure amorphous palladium. a) Total PDF, b) Pd-Pd partial PDF, c) Pd-Si partial PDF, d) Si-Si partial PDF.

The second coordination shell is located between 4.2 Å and 5.7 Å and is composed of two main peaks [3] with some reminiscence of multiple secondary peaks, as can be seen in the RDF of Figure 3 a); this complex second coordination shell was first reported by Fukunaga and Suzuki [16] and has been confirmed by other authors [23,24]. The partial Pd-Pd is the dominant factor in the bimodality of the second coordination shell; this bimodality is characteristic of all amorphous metals [3,25,26]. However, there seems to be secondary peaks present in some of the PdSi samples, these secondary peaks are due to the partial Pd-Si, as can be seen in Figure 3 b). Because of the complexity in the second coordination shell, the study of the partial PDFs is required to identify the origin of every peak [27,28]. In Table 1 the two more prominent peaks of the first coordination shell, as well as the three more prominent peaks in the second coordination shell, are reported. The first peak in the second coordination shell is due to the Pd-Pd partial and is located around 4.65 Å on average; the second peak is due to Pd-Si partial and on average it is located around 5.03 Å, and finally, the third peak is due to the Pd-Pd partial and is located at 5.32 Å on average for all samples studied in this work. Remarkably, the second coordination shell peaks do not have a clear dependence with the silicon concentration, in contrast with the first coordination peaks as previously discussed.

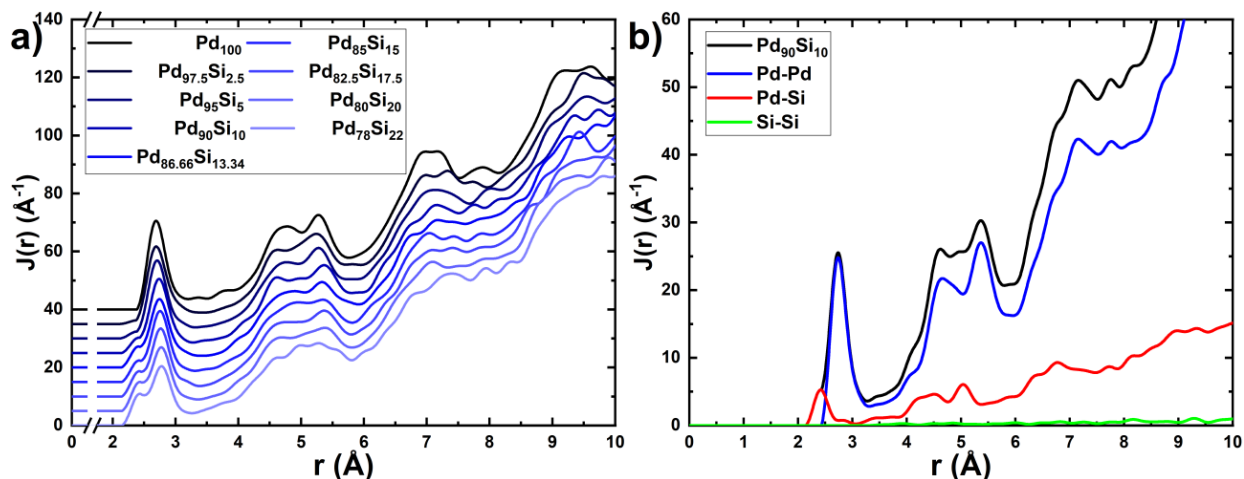

**Figure 3.** Radial Distribution Functions (RDFs) for amorphous PdSi alloys. a) Total RDFs for the eight concentrations of amorphous PdSi alloys and pure amorphous palladium. b) Partial RDFs for the $Pd_{90}Si_{10}$ sample.

**Table 1.** Peak positions for the first, $R_1$-, and second, $R_2$-, coordination shells in the total PDFs for the amorphous PdSi alloys and for the pure amorphous palladium.

| Alloy | $R_{1-1}$ Pd-Si | $R_{1-2}$ Pd-Pd | $R_{2-1}$ Pd-Pd | $R_{2-2}$ Pd-Si | $R_{2-3}$ Pd-Pd |
|---|---|---|---|---|---|
| $Pd_{78}Si_{22}$ | 2.425 | 2.765 | 4.595 | 4.995 | 5.355 |
| $Pd_{80}Si_{20}$ | 2.425 | 2.765 | 4.625 | 5.085 | 5.385 |
| $Pd_{82.5}Si_{17.5}$ | 2.415 | 2.755 | 4.685 | 5.045 | 5.375 |
| $Pd_{85}Si_{15}$ | 2.415 | 2.735 | 4.705 | 5.015 | 5.345 |
| $Pd_{86.66}Si_{13.34}$ | 2.405 | 2.745 | 4.675 | 5.005 | 5.325 |
| $Pd_{90}Si_{10}$ | 2.405 | 2.725 | 4.615 | 5.025 | 5.335 |
| $Pd_{95}Si_5$ | 2.385 | 2.695 | 4.695 | 5.055 | 5.255 |
| $Pd_{97.5}Si_{2.5}$ | 2.395 | 2.685 | 4.585 | 4.985 | 5.205 |
| $Pd_{100}$ | - | 2.685 | 4.655 | - | 5.255 |
| **Average** | 2.409 | 2.728 | 4.648 | 5.026 | 5.315 |

Coordination numbers, CNs, is another useful tool in the analysis of the structure of amorphous materials. However, determinations of CNs must be made delicately, because they are very sensitive to the cutoff of the interaction, and small errors in determining the first minimum of the pPDFs can yield noticeable differences in the results [8,22,24].

To calculate the CNs, two different procedures were used. The first procedure was by direct counting of bonded atoms, which is not trivial in metallic materials like palladium-rich PdSi. The Bond Calculation option included in the Materials Studio Visualizer, and also implemented in the **Correlation** software [18], was used. Two atoms are considered bonded if the distance between them is less than the sum of the covalent radii of the corresponding atoms times the bond parameter; in this work, a bond parameter of 1.2 was used. The second procedure was integrating the partial RDFs up to a cutoff radii, equal to the average minimum of every partial. To have a consistent result, the minimum after the first peak for every partial was determined, then an average of these minimums was taken over all studied samples, resulting in the cutoff radii of 3.32 Å for the Pd-Pd partial, 3.03 Å for the Pd-Si and the Si-Pd partial, and 3.06 Å for the Si-Si partial. These results can be seen in Table 2. The agreement between both procedures is remarkable and gives us confidence in our results for the CNs.

**Table 2.** Partial CNs calculated with the **Correlation** software using a bond parameter of 1.2 and with the integral of the partial RDFs, with a cutoff distance for the first peak of 3.32 Å for the Pd-Pd partial, 3.03 Å for the Pd-Si and the Si-Pd partial, and 3.06 Å for the Si-Si partial.

| Alloy | $Z_{Pd-Pd}$ | | $Z_{Pd-Si}$ | | $Z_{Si-Pd}$ | | $Z_{Si-Si}$ | |
|---|---|---|---|---|---|---|---|---|
| | [Correlation] | [Minimum] | [Correlation] | [Minimum] | [Correlation] | [Minimum] | [Correlation] | [Minimum] |
| $Pd_{78}Si_{22}$ | 9.63 | 9.64 | 2.39 | 2.42 | 8.38 | 8.46 | 0.33 | 0.38 |
| $Pd_{80}Si_{20}$ | 10.16 | 10.23 | 2.18 | 2.18 | 8.79 | 8.79 | 0.09 | 0.14 |
| $Pd_{82.5}Si_{17.5}$ | 10.43 | 10.56 | 1.87 | 1.86 | 8.74 | 8.72 | 0.00 | 0.12 |
| $Pd_{85}Si_{15}$ | 10.62 | 10.65 | 1.53 | 1.49 | 8.81 | 8.56 | 0.13 | 0.12 |
| $Pd_{86.66}Si_{13.34}$ | 10.92 | 10.79 | 1.34 | 1.35 | 8.62 | 8.68 | 0.07 | 0.08 |
| $Pd_{90}Si_{10}$ | 11.01 | 10.96 | 1.01 | 1.01 | 8.91 | 8.89 | 0.00 | 0.00 |
| $Pd_{95}Si_{5}$ | 11.40 | 11.11 | 0.49 | 0.49 | 9.18 | 9.09 | 0.00 | 0.00 |
| $Pd_{97.5}Si_{2.5}$ | 11.74 | 11.44 | 0.26 | 0.26 | 9.00 | 8.94 | 0.00 | 0.00 |
| $Pd_{100}$ | 11.92 | 11.45 | - | - | - | - | - | - |

The total silicon coordination goes from 8.7 for c = 22 at % to 9.2 for c = 5 at % using **Correlation**, which coincides with the results of Durandurdu [24]; similar results were obtained for $Pd_{84}Si_{16}$ by Sadoc and Dixmier [29] and for $Pd_{81}Ge_{19}$ by Pethes *et al.* [30]. In Figures 4 and 5 the information concerning the nearest neighbors and the FK polyhedra is shown. The most frequent FK polyhedra in amorphous PdSi alloys are represented in Figure 5 [31,32]. The first FK polyhedra is CN8 a Si-centered cluster with 8 first neighbors (Figure 5 a)), these polyhedra are predominant for c > 13.34 at %, and are the most important polyhedra for $Pd_{86.66}Si_{13.43}$; the Voronoi index for the CN8 FK polyhedra is <0 3 5 0 0> [33–35]. The second polyhedron is CN9 a Si-centered cluster with 9 first neighbors, as seen in Figure 5 b), these polyhedra are the most abundant in almost all concentrations studied, except for $Pd_{86.66}Si_{13.34}$ and $Pd_{95}Si_{5}$. The CN9 FK polyhedra have a corresponding Voronoi index of <0 3 6 0 0> in agreement with Durandurdu [24], the third Si-centered FK polyhedra are CN10 as seen in Figure 5 c), with a corresponding Voronoi index of <0 3 7 0 0> for the corresponding Voronoi cell. The fourth FK polyhedra are CN11, a Pd-centered cluster with 11 first neighbors, at least 8 of them being palladium atoms and at most 3 silicon atoms, as seen in Figure 5 d). The fifth FK polyhedra are CN12, a Pd-centered cluster with 12 first neighbors, at least 9 of them being palladium atoms and at most 3 of them being silicon atoms (Figure 5 e)), and finally CN13 a Pd-centered cluster with 13 first neighbors, at least 9 of them being palladium atoms and at most 4 of them being silicon atoms, see Figure 5 f). The total palladium coordination goes from 12.3 for c = 22 at % to 11.9 for pure palladium, in agreement with previous results [16,23,24,36].

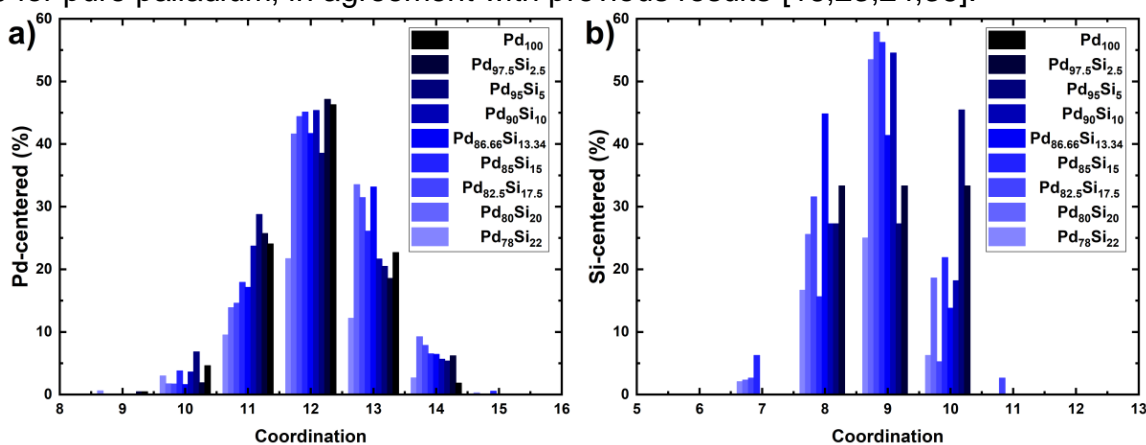

**Figure 4.** Total coordination numbers (CNs) for amorphous PdSi alloys and pure amorphous palladium. a) Pd-centered CNs, the most prominent Frank-Kasper polyhedron is CN12, b) Si-centered CNs, the most prominent Frank-Kasper polyhedron is CN9.

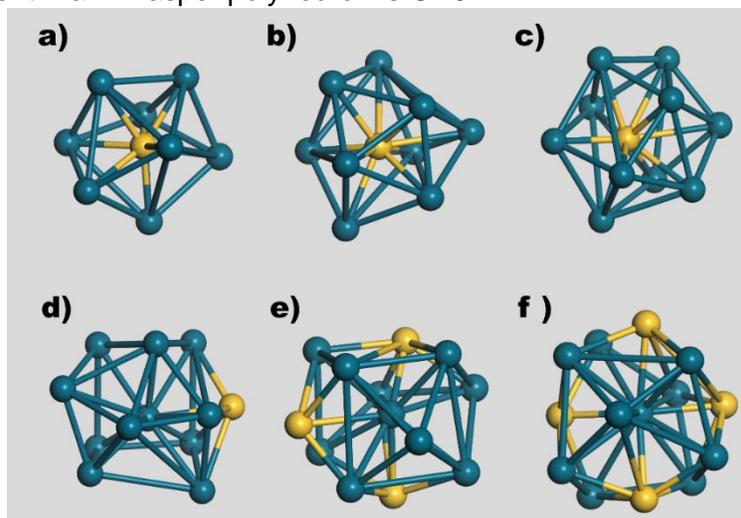

**Figure 5.** Frank-Kasper polyhedra for amorphous PdSi alloys and pure amorphous palladium. a) CN8 polyhedron, maximum presence in $Pd_{86.66}Si_{13.34}$, b) CN9 polyhedron is the most relevant Si-centered polyhedron, c) CN10 polyhedron relevant for c < 10 at %, d) CN11 polyhedron second most relevant Pd-centered polyhedron for c < 10 at %, e) CN12 polyhedron, more relevant Pd-centered polyhedron, f) CN13 polyhedron, more relevant polyhedra for c > 10 at % (Pd atoms in blue, Si atoms in yellow).

Two of the most prevalent polyhedra in amorphous PdSi alloys are CN12, which is the ideal condition in a perfect icosahedron, and CN10, which under ideal conditions is a Triaugmented triangular prism. Both polyhedra are part of the deltahedron, and the surface of the perfect polyhedra is constituted exclusively by equilateral triangles as reported by Frank and Kasper [31]. It is no surprise that the most prominent peak in the Plane Angle Distributions, PADs, is located around 60°, as can be seen in Figure 6 a). This peak in the total PAD is constituted by 3 different peaks in the ideal icosahedron, the surface triangles are equilateral, but the triangles that form with the center of the clusters are slightly deformed: 6 of them at 59.28° and 3 of them at 63.44° for every tetrahedral pyramid. This constitutes the icosahedral cluster shown in red lines in Figure 6 b).

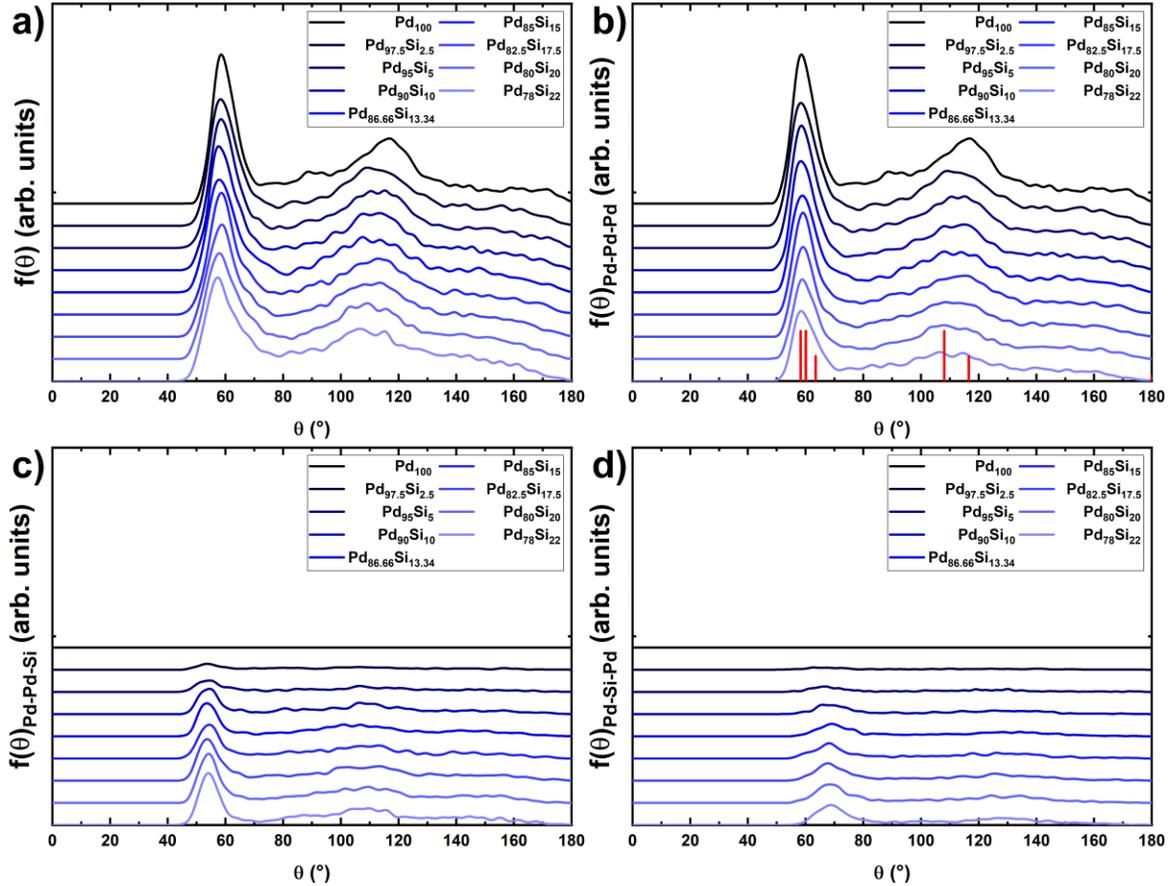

**Figure 6.** Plane Angle Distributions for the amorphous PdSi alloys and pure amorphous palladium, a) Total PADs for amorphous PdSi alloys, b) Pd-Pd-Pd partial PADs, perfect icosahedral PAD in red lines. c) Pd-Pd-Si partial PADs and d) Pd-Si-Pd partial PADs.

While the perfect icosahedral has only 5-fold symmetry in all its vertices, all other FK polyhedra present in amorphous PdSi have other symmetries, the second most prevalent symmetry is the 4-fold; and it is most prevalent in all Si-centered clusters (CN8, CN9, and CN10), This explains the peaks in 89.2° and 90.6° in the Pd-Pd-Pd partial PADs, at 90° the peak is not exactly single because the square-like structures are deformed in the amorphous. The most prominent peak in the partial Pd-Pd-Si PADs is at 55° (Figure 6 c)). Additionally, the most prominent peak in the partial Pd-Si-Pd PADs is at 69° (Figure 6 d)); These peaks differ from the 60° of the equilateral triangles due to the shorter distance in the Pd-Si bonds than in the Pd-Pd bonds, forming isosceles triangles instead of the partial equilateral ones in the Pd-Pd-Pd PADs.

The CN13 cluster becomes a very important FK polyhedron for concentrations c > 15 at %. However, this is a cluster that should not exist in the hard sphere approximation [37,38], its presence normally requires that one extra atom gets trapped between the first and second coordination shells. This explains the non-zero valley between the first and second neighbors, as can be seen in simulations and experiments (Figure 1). In the amorphous PdSi alloys as well as other metal-metalloid alloys the difference in covalent radii makes it possible to insert an extra atom in the first coordination shell. For this reason, there is an increase in the total Pd CNs from 12 to 12.35 for $Pd_{82.5}Si_{17.5}$, as can be seen in Figure 5 f).

A byproduct of this study is that there seems to be sudden structural rearrangements at $Pd_{86.66}Si_{13.34}$, since there is a change in the Si-centered FK polyhedra from CN9 to CN8. This, in conjunction with other factors like the appearance of Si-Si first neighbors and the change in magnetic properties from a paramagnetic to a diamagnetic alloy [1,39], may indicate the presence of a paramagnetic-diamagnetic transition due to the formation of a Si-Si bond. More studies are needed to clarify this point.

Finally, in Figures 7 a representation of the 8 amorphous cells generated in this work, is given for completeness. The pure palladium amorphous cell is reported in Ref. 3.

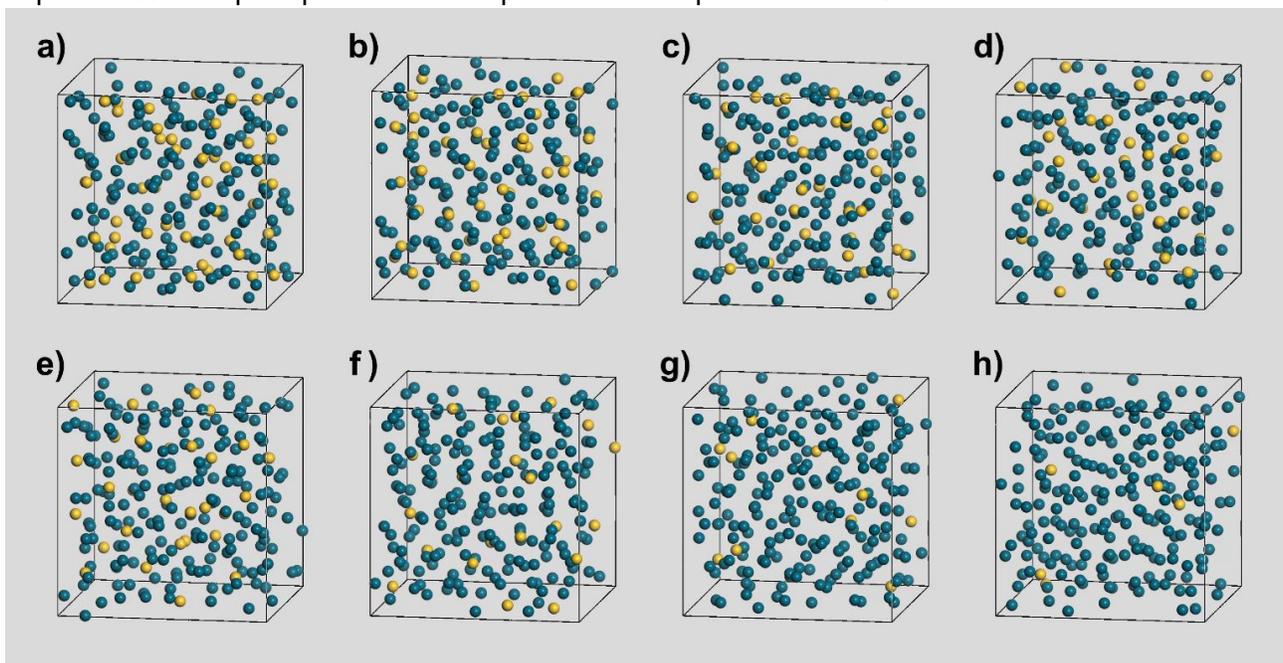

**Figure 7.** Atomic structures of the amorphous PdSi cells for the 8 concentrations studied in this work. a) $Pd_{78}Si_{22}$, b) $Pd_{80}Si_{20}$, c) $Pd_{82.5}Si_{17.5}$, d) $Pd_{85}Si_{15}$, e) $Pd_{86.66}Si_{13.34}$, f) $Pd_{90}Si_{10}$, g) $Pd_{95}Si_{5}$, h) $Pd_{97.5}Si_{2.5}$ (Pd atoms in blue, Si atoms in yellow).

**Conclusions**

We generated 8 samples of amorphous $Pd_{100-c}Si_c$ alloys with concentrations c = 2.5, 5, 10, 13.34, 15, 17.5, 20 and 22 at % with the *undermelt-quench* approach developed in our group, an *ab initio* procedure driven by molecular dynamics. The resulting atomic structures were analyzed by the pair distribution functions, the reduced pair distribution functions, the radial distribution functions and the plane angle distribution functions, finding the structures with a local minimum energy that represent the alloy topology. Possible structural changes around 13.34 at %, which may be related to the magnetic transition found by our workgroup [1] in the amorphous solid alloys, as well as by Müller *et al.* in the liquid alloys [39], are reported.

The palladium average coordination number for c > 13.34 at % is 12.2, in agreement with previous studies for amorphous PdSi, while it becomes 12 for c < 13.34 at % for pure amorphous palladium. We found that the most representative atomic structures centered on Pd are the CN12 FK polyhedron, which corresponds to deformed icosahedra, while Si-centered clusters are represented by a FK CN9 cluster, a deformed triaugmented triangular prism; both of these structures are convex polyhedra constituted by equilateral triangles.

## Graphical Abstract

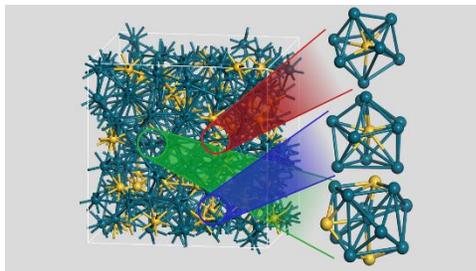

### Acknowledgements
Isaías Rodríguez thanks PAPIIT, DGAPA-UNAM, for his postdoctoral fellowship. David Hinojosa-Romero acknowledges Consejo Nacional de Ciencia y Tecnología (CONACyT) for supporting his graduate studies. Ariel A. Valladares, Renela M. Valladares and Alexander Valladares thank DGAPA-UNAM (PAPIIT) for continued financial support to carry out research projects under Grants No. IN104617 and IN116520. M.T. Vázquez and O. Jiménez provided the information requested. A. Lopez and A. Pompa assisted with the technical support and maintenance of the computing unit at IIM-UNAM. Simulations were partially carried at the Computing Center of DGTIC-UNAM under the project LANCAD-UNAM-DGTIC-131.

### Author contributions
A.A.V., A.V. and R.M.V. conceived this research and designed it with the participation of I.R. and D.H.-R. I.R. did all the simulations. All authors discussed and analyzed the results. A.A.V. wrote the first draft and the other authors enriched the manuscript.

### Competing interest
The authors declare no competing interests.

### Data and code availability
The datasets generated during this study are available from the corresponding author on reasonable request. Correlation is a code that can be accessed through GitHub repository:
https://github.com/Isurwars/Correlation

### Additional information
Correspondence and requests for materials should be addressed to A.A.V.